\documentclass[aps,prb,amsmath,amssymb,amsfont,showpacs,superscriptaddress,reprint]{revtex4-1}

\usepackage{graphicx}
\usepackage[usenames,dvips]{color}
\usepackage{times}
\usepackage{natbib}
\usepackage{subfigure}
\usepackage{soul}

\usepackage{xr}

\usepackage{bm}

\begin{document}

\newcommand{\dd}[1]{\mathrm{d}#1}
\newcommand{\kb}{k_\text{B}}
\newcommand{\Td}{T_\text{D}}
\newcommand{\comment}[2]{#2}
\newcommand{\todo}[1]{{\color{red}#1}}

\newcommand{\JO}[1]{\color{blue}#1}
\newcommand{\jav}[1]{\textcolor{red}{#1}}

\title{Spin-relaxation time in materials with broken inversion symmetry and large spin-orbit coupling}

\author{L\'enard Szolnoki}
\affiliation{Department of Physics, Budapest University of Technology and Economics and
MTA-BME Lend\"{u}let Spintronics Research Group (PROSPIN), POBox 91, H-1521 Budapest, Hungary}

\author{Annam\'aria Kiss}
\affiliation{Institute for Solid State Physics and Optics, Wigner Research Centre for Physics, Hungarian Academy of Sciences, POBox 49, H-1525 Budapest, Hungary}
\affiliation{MTA-BME Lend\"{u}let Spintronics Research Group (PROSPIN), POBox 91, H-1521 Budapest, Hungary}

\author{Bal\'azs D\'ora}
\affiliation{Department of Theoretical Physics, Budapest University of Technology and Economics and
MTA-BME Lend\"{u}let Spintronics Research Group (PROSPIN), POBox 91, H-1521 Budapest, Hungary}

\author{Ferenc Simon}
\email[Corresponding author: ]{f.simon@eik.bme.hu}
\affiliation{Department of Physics, Budapest University of Technology and Economics and
MTA-BME Lend\"{u}let Spintronics Research Group (PROSPIN), POBox 91, H-1521 Budapest, Hungary}

\pacs{76.30.Pk, 71.70.Ej, 75.76.+j	}

\date{\today}

\begin{abstract}
We study the spin-relaxation time in materials where a large spin-orbit coupling (SOC) is present which breaks the spatial inversion symmetry. 
Such a spin-orbit coupling is realized in zincblende structures and heterostructures with a transversal electric field and the spin relaxation is usually 
described by the so-called D'yakonov-Perel' (DP) mechanism. We combine a Monte Carlo method and diagrammatic calculation based approaches in our study; the former 
tracks the time evolution of electron spins in a quasiparticle dynamics simulation in the presence of the built-in spin-orbit magnetic fields and the latter builds on the spin-diffusion propagator by Burkov and Balents [Burkov and Balents Phys. Rev. B. \textbf{69}, 245312 (2004).]. 
Remarkably, we find a parameter free \emph{quantitative} agreement between the two approaches and it also returns the conventional result of the DP mechanism 
in the appropriate limit. We discuss the full phase space of spin relaxation as a function of SOC strength, its distribution, and the magnitude of the momentum 
relaxation rate. This allows us to identify two novel spin-relaxation regimes; where spin relaxation is strongly non-exponential and the spin relaxation equals 
the momentum relaxation. A compelling analogy between the spin-relaxation theory and the NMR motional narrowing is highlighted.
\end{abstract}

\maketitle

\section*{Introduction}

It is an intriguing possibility to employ the electron spins as information carriers (known as spintronics \cite{FabianRMP}). This prospect has revived the 
experimental and theoretical studies of spin-relaxation in semiconductors and metals. It is the spin-relaxation time which characterizes how rapidly a non-equilibrium 
spin population decays and it therefore determines whether a material is suitable for spintronics purposes. The theory of spin-relaxation differs for materials with 
and without spatial inversion symmetry \cite{FabianRMP,WuReview}: the Elliott-Yafet (EY) theory \cite{Elliott,yafet1963g} describes the former while the so-called 
D'yakonov-Perel' (DP) mechanism describes the spin-relaxation for the latter case \cite{DyakonovPerelSPSS1972}. We note that the common physical picture to unify 
the two approaches was developed in Ref. \onlinecite{Boross2013}. 

The DP theory describes the dominant spin-relaxation mechanism in large band-gap III-V semiconductors (e.g. GaAs) with the zincblende structure (the so-called bulk 
inversion asymmetry) and for semiconductor heterostructures with an applied transversal electric field (the so-called structure inversion asymmetry). These two cases 
are known as Dresselhaus or Bychkov-Rashba spin-orbit coupling (SOC), respectively. 

The DP spin-relaxation mechanism turned out to be particularly relevant for novel, spintronics candidate materials with two-dimensional structure, such as e.g. 
mono or bi-layer graphene \cite{GeimRMP,HuertasPRL2009} and transition metal dichalcogenide monolayers \cite{Heinz_TMDC,Wu_TMDC,Ozyilmaz_TMDC_DP}.

\begin{figure}[htp]
\begin{center}
\includegraphics[scale=0.4]{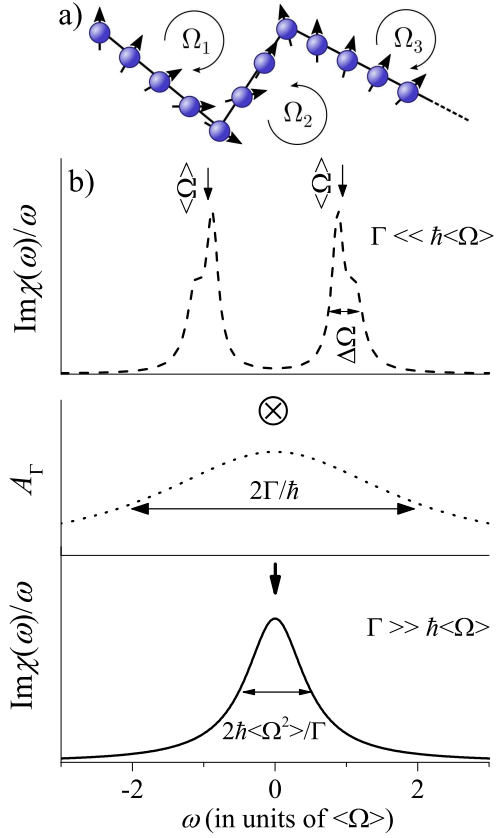}
\caption{Schematics of the conventional D'yakonov-Perel' spin-relaxation mechanism (a): the electron spin precesses around the SOC fields and scatter on a $\tau$ timescale. Each scattering results in a new random $\bm{k}$ value and a different vector of the SOC field. (b) In the absence of momentum scattering, i.e. the clean limit, the dynamic spin-susceptibility, $\chi(\omega)$, is given by the distribution of the SOC fields. The quasi-particle spectral function, $A_{\Gamma}$, has a width of $2\Gamma/\hbar$ and the finite momentum lifetime gives rise to the "dirty" case. We show $\text{Im}\chi(\omega)/\omega$ for clarity as $\text{Im}\chi(\omega)$ is an odd function.}
\label{fig1_DP_schematics}
\end{center}
\end{figure}

The spin-relaxation mechanism, which dominates in materials without inversion symmetry, is depicted schematically in Fig. \ref{fig1_DP_schematics}a. The inversion 
symmetry breaking SOC splits the spin-up/down states at the Fermi level and acts as a $\bm{k}$ dependent magnetic field or SOC field. When the electrons are treated 
in the quasi-particle approximation, they are assumed to move around in the material and suffer momentum scattering on a timescale of $\tau$. The electron spins precess 
around the axis of the SOC field with a corresponding Larmor frequency, $\Omega(\bm{k})$. The rigorous derivation of the DP result \cite{DyakonovPerelSPSS1972,FabianRMP} 
gives that the spin-relaxation time, $\tau_{\text{s}}$ is inversely proportional to $\tau$ if the $\left<\Omega\right>\cdot \tau \ll 1$ holds (here $\left<\Omega\right>$ 
is an average value of the Larmor frequencies). When rewritten for the quasiparticle momentum scattering rate: $\Gamma=\hbar/\tau$ and the 
 spin scattering rate: $\Gamma_{\text{s}}=\hbar/\tau_{\text{s}}$, the DP result reads:

\begin{gather}
\Gamma_{\text{s}}=\alpha \frac{\left<\left|\mathcal{L}\right|^2\right>}{\Gamma}
\end{gather}
where $\alpha$ is a band structure dependent parameter around unity and $\left<\mathcal{L}\right>$ is an average value of the SOC induced splitting and is related to the SOC fields by $\left<\mathcal{L}\right>=\hbar \left<\Omega\right>$. 

Rather than a rigorous derivation, we give an illustration of the DP result in Fig. \ref{fig1_DP_schematics}b. In the clean limit (i.e. $\Gamma=0$), 
the dynamic spin-susceptibility, Im$\chi(\omega)$, is an odd function around the origin which is 
given solely by the distribution of the SOC fields and it describes pure dephasing without relaxation. Im$\chi(\omega)$ could be 
observed in a clean semiconductor by e.g. an electron spin resonance experiment. The conventional DP theory applies in the dirty limit, i.e. 
when $\Gamma \gg \hbar\left<\Omega\right>$, which is depicted in Fig. \ref{fig1_DP_schematics}b. (dotted curve is the quasi-particle spectral function). 
This is in fact the textbook situation of motional narrowing, which is well known in NMR spectroscopy \cite{AbragamBook} and it formally leads to the DP 
result. In the spintronics literature, the DP regime is also known as \emph{motional slowing} of the spin-relaxation.

The opposite limit, i.e. when $\Gamma\lesssim\hbar\left<\Omega\right>$ is less well-known and studied although recent experiments on ultra-pure GaAs 
indicates \cite{Oscillatory_DP_exp1,Oscillatory_DP_exp2} that this regime is accessible. In addition, the theoretical foundations of this regime were 
presented in Ref. \onlinecite{Oscillatory_DP_theor}. This regime corresponds to semiconductors with a large SOC and low quasiparticle scattering. 
The $\Gamma \ll \hbar\left<\Omega\right>  $ limit is discussed in Ref. \onlinecite{FabianRMP} and it is argued that the distribution of the SOC fields 
causes a rapid dephasing of a spin ensemble orientation on the timescale of $\tau_{\text{s}}=1/\Delta \Omega$, where the latter quantity is the 
characteristic distribution width or variance of the SOC fields. This process is analogous to the so-called reversible dephasing in NMR spectroscopy with the timescale known as $T_2^{*}$ (Ref. \onlinecite{SlichterBook}). 
A momentum scattering after the dephasing causes memory loss therefore the observable spin-decay time roughly equals the spin-dephasing time. In our notation, the spin-relaxation rate was suggested to read in this limit as \cite{FabianRMP}:

\begin{gather}
\Gamma_{\text{s}}=\Delta \mathcal{L}
\label{Fabian_result}
\end{gather}
where $\Delta \mathcal{L}=\hbar \Delta \Omega$ is the distribution width of the SOC splitting. 

While we believe this qualitative picture to be correct, this regime requires a more quantitative 
description and as we show herein, the simple description breaks down depending on the relative 
magnitude of $\Gamma$ and $\Delta \mathcal{L}$. Another unresolved issue is whether the motional
 narrowing description of D'yakonov and Perel' could be continued for cases when its conditions are not fulfilled. This regime would be relevant for clean semiconductors.

These shortcomings motivated us to study the full phase space of $\left(\Gamma,\left<\mathcal{L}\right>,\Delta \mathcal{L}\right)$ 
with emphasis on the case of large SOC, i.e. when $\Gamma \ll \mathcal{L}$. We compare two distinct approaches: a Monte Carlo method 
which is essentially a quasiparticle kinetics based approach using numerical simulation and the diagrammatic technique, accounting for both SOC and impurity scattering. We studied the Bychkov-Rashba Hamiltonian for a two-dimensional electron gas with both methods. The two approaches 
yield \emph{quantitatively} identical results for the spin-relaxation time without adjustable parameters. We observe a non 
single-exponential spin decay with both methods in certain cases. This validates the Monte Carlo approach and lets us present a 
method which allows calculation of $\tau_{\text{s}}$ (or $\Gamma_{\text{s}}$) for an \emph{arbitrary} SOC model including e.g. 
the Dresselhaus SOC. We identify a yet unknown regime: when the distribution of the SOC is sharper than the broadening parameter, 
i.e. $\Gamma \gg \Delta \mathcal{L}$, then $\Gamma_{\text{s}}\approx\Gamma$ is realized. For the case of $\Gamma \lesssim \left(\left<\mathcal{L}\right>, \Delta \mathcal{L}\right)$ 
a strongly non-exponential spin decay is observed. All the results can be elegantly visualized by considering the 
evolution of the dynamic spin-susceptibility from the clean to the dirty limit. We note that we focus on the effect of 
inversion symmetry breaking SOC fields on spin-relaxation and we do not discuss other mechanisms, such as e.g. the Elliott-Yafet mechanism \cite{Elliott,yafet1963g}. 

\section*{Methods}
\subsection*{The Monte Carlo simulations}

The time evolution of the electron spin direction in the presence of internal spin-orbit coupling is studied with a 
Monte Carlo approach which essentially mimics the mathematical description of D'yakonov and Perel' \cite{DyakonovPerelSPSS1972}: 
an initially polarized electron spin ensemble travels in a solid where the SOC related and momentum ($\bm{k}$) dependent magnetic 
fields ($\bm{B}(\bm{k})$) are present. The spins precess freely around the SOC fields as classical variables with angular momentum 
$\bm{\Omega}(\bm{k})$ between two scattering events, where $\bm{k}$ points to the Fermi surface. The scattering, what we model as a stochastic process,  
induces a new random $\bm{k}$ on the Fermi surface thus precession starts around a new axis. The momentum-relaxation is described by a Poisson process with 
an expectation value of $1/\tau$. Thus the time between successive scattering events follows an exponential distribution with the same $1/\tau$ parameter. 
We also assume that the momentum scatters uniformly on the Fermi surface and that the scattering does not affect the spin direction, i.e. the Elliott-Yafet type spin-flip mechanisms \cite{Elliott,yafet1963g} are not considered. 

After running the simulation on several independent spins, the mean spin component is sampled in uniform time intervals.
We keep the momentum relaxation $\tau=1$ constant in the simulations and vary the strength of the spin-orbit coupling and the proper time dependent data are obtained by a rescaling. The spin is measured in units of $\hbar/2$.

\begin{figure}[htp]
\begin{center}
\includegraphics[scale=0.4]{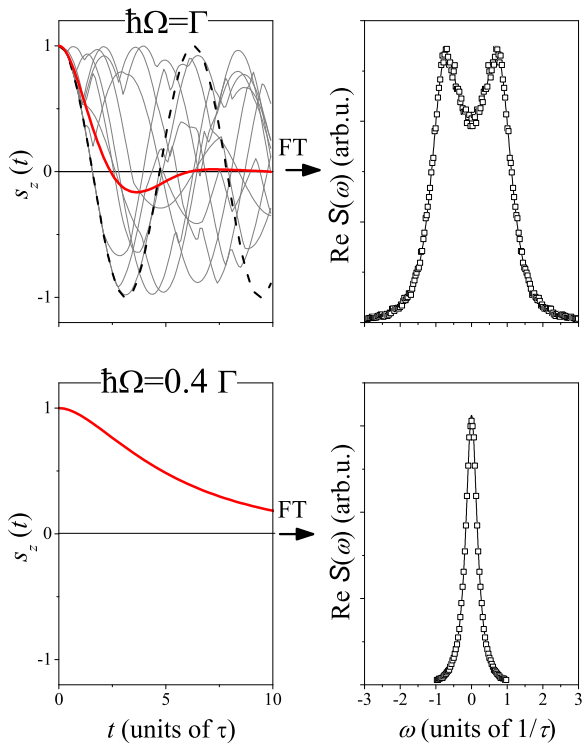}
\caption{Simulation of the time dependent magnetization for an ensemble of electrons under the Bychkov-Rashba SOC Hamiltonian for $\hbar \Omega=\Gamma$ 
(upper panel) and $\hbar\Omega=0.4\Gamma$ (lower panel). Thick solid curve in the upper left image is the averaged $s_{\text{z}}$, thin solid curves are 
the magnetizations of a few individual electrons. Dashed curve show the oscillating magnetization in the absence of momentum scattering. The real part 
of the Fourier transform of $s_z(t)$, $\text{Re}\mathcal{S}(\omega)$ (symbols) and Lorentzian fits (solid curves) as explained in the text, are also shown.}
\label{fig2_MC_basics}
\end{center}
\end{figure}

{The physical picture behind this approach is that i) the system is at $T=0\,\text{K}$, ii) $\Gamma$ appears solely through the momentum scattering time without considering its origin (i.e. impurities, phonons etc)
and the $\Gamma$ related state broadening,  iii) we consider only $k=k_{\text{F}}$ states, i.e. the SOC induced band splitting is disregarded in the initial state, and that iv) all electrons on the Fermi surface are spin polarized and the rest of the Fermi sea
is unpolarized in the initial state. Neglecting the effect of $\Gamma$ on the band structure for the Monte Carlo approach is justified by comparing the result with that of the 
diagrammatic technique (which correctly accounts for the finite life-time effects) in Ref. \onlinecite{BurkovBalents}. Considering $k=k_{\text{F}}$ states only for the band structure without SOC splitting is a common approach in similar spin dynamics \cite
{BurkovBalents} studies and when calculating the SOC related dynamic spin-susceptibility \cite{ErlingssonPRB}. 
It essentially corresponds to neglecting corrections on the order of $\mathcal{L}/E_{\text{F}}$, where $E_{\text{F}}$ 
is the Fermi energy and further details about this approximation are provided in the Supplementary Material. 
The assumption iv) corresponds to the application of a small magnetic field for the time $t<0$ which causes all spins on the Fermi surface to be polarized and the rest of the Fermi sea to remain unpolarized and the magnitude of the required magnetic field is discussed below. 

Typical time dependent magnetization curve for an ensemble of electrons is shown in Fig. \ref{fig2_MC_basics}. In this case, we assumed a two-dimensional electron gas with a Bychkov-Rashba type SOC ($x$ and $y$ denotes coordinates within the plane): 

\begin{equation}
  H_0 = \frac{\hbar^2\bm{k}^2}{2m}
           + \frac{\mathcal{L}}{k_\text{F}}(s_x k_y - s_y k_x)
\label{2D_BR_Hamiltonian}					
\end{equation}
where the first term is the kinetic energy, $k_\text{F}$ is the Fermi wavenumber, $\bm{k}^2=k_x^2+k_y^2$, $s_{x,y}$ and $k_{x,y}$ are the components of the spin and momentum, 
respectively. The corresponding SOC related Larmor frequencies read: $\bm{\Omega}(\bm{k}) =\frac{\mathcal{L}}{\hbar k_{\text{F}}}\left[k_y,-k_x,0\right]$. We also assumed that 
the spins only on the Fermi surface are fully polarized along $z$ at $t=0$. We considered two cases, a large SOC ($\hbar\Omega=\Gamma$) and moderate SOC ($\hbar\Omega=0.4\Gamma$) and the result is shown in Fig. \ref{fig2_MC_basics}. 
As we show below, these are archetypes of the different relaxation regimes. A decaying magnetization is observed for both cases with and without an oscillating component. 
The real part of the Fourier transform of the time dependent $s_z(t)$ data, $\text{Re}\mathcal{S}(\omega)$ is also shown in Fig. \ref{fig2_MC_basics}, 
which depicts better the presence of the oscillation (peaks at $\omega \neq 0$) and a single decay (a nearly Lorentzian peak at $\omega =0$). The details of these 
calculations are analyzed below by fitting two Lorentzian curves to them, whose position and width give the frequency of the damped oscillations and the damping, 
respectively. These parameters are compared to the analytic calculations for the same Bychkov-Rashba Hamiltonian performed in Ref. \onlinecite{BurkovBalents}. 
The clean limit, i.e. $\Gamma=0$ would give a $\text{Re}\mathcal{S}(\omega)$ with two Dirac-delta peaks at $\omega=\pm \Omega$ upon neglecting terms of the order $\mathcal{L}/E_{\text{F}}$.

{We briefly discuss the relationship between $\mathcal{S}(\omega)$ and the dynamic spin-susceptibility, $\chi(\omega)$. The time dependent net magnetic moment, $s_z(t)$ is proportional to the sample magnetization. 
Assuming linear response theory to apply, the magnetization of the sample, $M$ is related to an external magnetic field, $B$ by:
\begin{gather}
s_z(t)\propto\mu_0 M(t)=\int_{-\infty}^t \chi\left(t-t'\right)B\left( t' \right) \text{d} t'
\label{chi_equation}
\end{gather}
where $\mu_0$ is the vacuum permeability. Our calculation assumes that the strength of the magnetic field is such that electrons on the Fermi surface are spin-polarized and electrons on the next level are already compensated by spin-degeneracy. 
The corresponding magnetization is given by: $M\approx \frac{g \mu_{\text{B}}}{V_{\text{c}}}\frac{\Delta k}{k_{\text{F}}}$, where $\Delta k$ denotes the typical distance between allowed momentum space points.
A magnetic field of $g \mu_{\text{B}}B_0=\frac{\Delta k}{k_{\text{F}} g(E_{\text{F}})}$ (where $g(E_{\text{F}})$ is the density of states at the Fermi energy) is required to sustain such a magnetization at $T=0$. 
This magnetic field tends to zero in the thermodynamic limit, i.e. it is sufficiently small for the linear response theory to be valid. 

Considering that in our model $B(t')=B_0\Theta(-t')$ (where $\Theta(t')$ is the Heaviside function), i.e. it is switched off at $t=0$, a straightforward calculation leads to 
$i\omega\cdot\mathcal{S}(\omega) \propto \chi(\omega)$ the proportionality constant being the infinitesimal magnetic field. 
We numerically confirm this proportionality 
for the two-dimensional Bychkov-Rashba and Dresselhaus Hamiltonians without impurities (data and additional discussion is given in the Supplementary Material) i.e. that 
$\omega\cdot\text{Re}\mathcal{S}(\omega) \propto \text{Im}\chi(\omega)$ holds in the clean limit. We used $\chi(\omega)$ data published in 
Ref. \onlinecite{ErlingssonPRB} for the comparison within the above discussed $\mathcal{O}\left(\mathcal{L}/E_{\text{F}}\right)$ approximation. 
From Eq. \eqref{chi_equation}, the above relation is expected to remain also valid in the dirty limit, i.e. 
 $\omega\cdot\text{Re}\mathcal{S}(\omega) \propto \text{Im}\chi(\omega)$. 
We note that since $\text{Im}\chi(\omega)$ is an odd function of frequency, $\text{Re}\mathcal{S}(\omega)$ must be even.
In general, 
$\text{Im}\chi(\omega)$ is known to carry information about the spin dynamics and losses, which explains why we present the numerically obtained $\text{Re}\mathcal{S}(\omega)$ throughout this contribution.

The Monte Carlo method we use  can be readily applied for an arbitrary distribution of the SOC fields, e.g. for the three-dimensional Dresselhaus Hamiltonian as we discuss below. 
The system specific parameters are present through the actual $\bm{\Omega}(\bm{k})$ function. We note that in general the spin-relaxation is anisotropic, i.e. different oscillations 
and damping are observed for the same SOC distribution depending on the starting polarization direction of the spins.

\subsection*{Results of the diagrammatic technique}

  Burkov and Balents calculated the spin-relaxation in a two-dimensional electron gas \cite{BurkovBalents} for a Rashba type SOC for an arbitrary value of the magnetic field. 
Their calculation turns out to be very general and although they applied it in the D'yakonov-Perel' regime (i.e. for weak SOC), it can be also used for an arbitrary strength of the SOC. The model Hamiltonian was the same as Eq. \eqref{2D_BR_Hamiltonian}, i.e. the 2DEG with Rashba SOC. The so-called 
spin-diffusion propagator, $D(\omega)$ was calculated in Ref. \onlinecite{BurkovBalents} with a diagrammatic technique. We use Eqs.~36 and Eq.~38 in Ref. \onlinecite{BurkovBalents} 
to obtain the spin-relaxation times from the spin-diffusion propagators, which reads in the $z$ direction as:

  \begin{equation}
    D_{zz}(\omega) = \frac{ (-i \Gamma +\mathcal{L}-\hbar\omega )
                    (i \Gamma +\mathcal{L}+\hbar\omega )
                  }
                  { -\hbar^2\omega^2 
                    -i \Gamma \hbar\omega 
                    + \mathcal{L}^2
                  }.
									\label{BB_spindiffusion_propagator}
  \end{equation}

The real and imaginary parts of the two poles ($\omega_{1,2}$) of Eq. \eqref{BB_spindiffusion_propagator} describe the oscillation frequency and the damping of the spin-propagation. The poles read:

  \begin{equation}
    \hbar\omega_{1,2}= \frac{ -i \Gamma\pm \sqrt{4 \mathcal{L}^2 - \Gamma^2}}{2}.
  \end{equation}
The spin-relaxation time is obtained from the poles as: 
  \begin{equation}
    \frac{1}{\tau_\text{s}}=-\text{Im}\,\omega_{1,2}.
  \end{equation}
Similar results can be obtained for the diffusion propagator which is perpendicular to the quantization axis ($D_{xy}$) and the poles are the roots of a third order polynomial:

\begin{equation}
	(\hbar\omega)^3
	+ 2 i \Gamma (\hbar\omega)^2
	- \left( \Gamma^2 + \mathcal{L}^2 \right)\hbar\omega
	- i \frac{ \mathcal{L}^2 \Gamma }{2}
	  = 0.
\end{equation}

We emphasize that the result of Burkov and Balents is valid for \emph{any} value of $\mathcal{L}$ and $\Gamma$ for the studied Hamiltonian. We use this result for comparison with the Monte Carlo simulations on the same Bychkov-Rashba SOC model system. We presume that analytic results could be obtained for other simple models of the SOC using the formalism of Ref. \onlinecite{BurkovBalents} but these are beyond our scope.

\section*{Results and discussion}

\subsection*{Validation of the Monte Carlo approach}

\begin{figure}[htp]
\begin{center}
\includegraphics[scale=0.4]{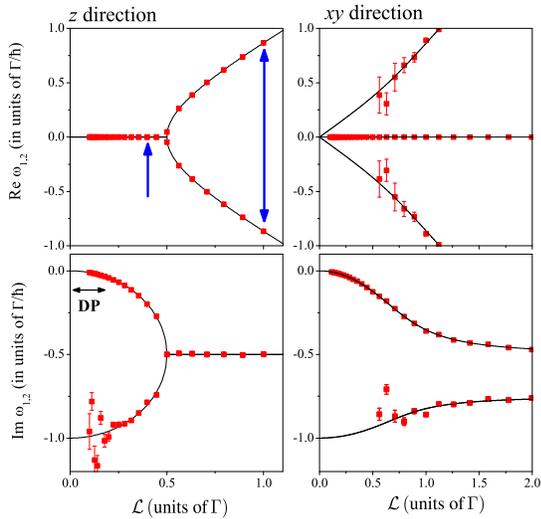}
\caption{Spin-relaxation parameters as obtained from the analytic calculation in Ref. \onlinecite{BurkovBalents} (solid lines) and the Monte Carlo simulations (symbols) for the two orientations of the initial spin polarization. The real part is the frequency of the damped oscillation and the imaginary part is the spin-relaxation rate in frequency units. Vertical blue arrows show the two cases which are discussed in the previous figure. Note that there is no scaling parameter between the two kinds of data. Horizontal arrow shows the conventional DP regime. Note the different horizontal scale for the $xy$ direction data. Scattering of some parameter values is a sign of a less reliable fit due to a vanishing spectral weight of the corresponding Lorentzian.}
\label{fig3_validation}
\end{center}
\end{figure}

As a first step, we validate the above described Monte Carlo approach by comparing the numerical results with that of the analytic calculations. The comparison is shown in Fig. \ref{fig3_validation}. The analytic result in Ref. \onlinecite{BurkovBalents} yields two parameters: the real and imaginary parts of poles of the spin-diffusion propagator ($\omega_{1,2}$) which are shown with solid curves in Fig. \ref{fig3_validation}. The earlier represents the frequency of the oscillating component, whereas the latter describes the damping or relaxation and it is the spin-relaxation rate in frequency units. A fit to the numerically obtained $\text{Re}\mathcal{S}(\omega)$ data with several Lorentzian components (including both Kramer-Kronig pairs) yield the position of the Lorentzian as well as its width. This result is shown in Fig. \ref{fig3_validation}. with symbols. We observe a surprisingly good agreement between the two types of data, which as we note is obtained without any \emph{adjustable paramaters}. Essentially, the two types of calculations consider the same SOC Hamiltonian (the Bychkov-Rashba) and a two-dimensional electron gas and the same approximation (neglect of SOC splitting of the band structure, i.e. zeroth order in $\mathcal{L}/E_{\text{F}}$) but the methods are quite different. We can omit the $\left<...\right>$ notation for the Bychkov-Rashba model as the SOC field consists of two delta functions when the $\mathcal{O}\left(\mathcal{L}/E_{\text{F}}\right)$ corrections are neglected.

A horizontal arrow in Fig. \ref{fig3_validation}. indicates the regime where the conventional DP mechanism is realized and both the numerical and Monte Carlo methods give $\Gamma_{\text{s}}=\mathcal{L}^2/\Gamma$ (i.e. herein $\alpha=1$). However, the figure also indicates the presence of two additional, previously unknown spin-relaxation regimes: on the far right, when $\Gamma \lesssim \mathcal{L}$, two peaks are present in with a broadening of $\Gamma_{\text{s}}=\alpha\Gamma$ (where $\alpha$=0.5). Another spin-relaxation regime occurs right below the "bifurcation point" ($\mathcal{L}=0.5 \Gamma$): therein two relaxation components with different broadening are present, i.e. it represents a non single-exponential relaxation. This regime crosses over smoothly to the conventional DP regime where a single exponential is observed: the weight of the component with larger broadening gradually disappears, which is apparent from the scatter in the parameters which are obtained by fitting the Monte Carlo data. 

The conventional DP mechanism is often referred to as a motional narrowing effect, which gives rise to the $\Gamma_{\text{s}}\propto \Gamma^{-1}$ behavior. In fact, motional narrowing is the extremal case of a theory known as "\emph{the effect of motion on the spectral lines}", which is well developed in e.g. NMR spectroscopy \cite{AbragamBook}. We show in the Supplementary Material that a conventional textbook description of the so-called two-site moving nuclei problem (after Ref. \onlinecite{AbragamBook}, Chapter X.) gives \emph{numerically identical} results to the parameters of the above described Bychkov-Rashba model for the full motional range. While this is an interesting analogy, it is a straightforward argument that the two situations, i.e. momentum scattering of an electron between different $\bm{\Omega}(\bm{k})$ and motion of nuclei between different sites with different Larmor frequencies, leads to the same result. A more interesting consequence of this analogy is that the full phase space of $\left(\Gamma,\mathcal{L}\right)$ can be regarded as a motional problem and it is not restricted to the DP regime.

\subsection*{Spin-relaxation for the Dresselhaus spin-orbit coupling}

The agreement between the spin-relaxation parameters as obtained from the Monte Carlo and from the analytic calculations validates the use of the numerical method for the two-dimensional electron gas with the Bychkov-Rashba SOC. Although it represents no formal proof, we believe that it justifies the use of the Monte Carlo method to obtain spin-relaxation parameters and eventually $\text{Re}\mathcal{S}(\omega)$ for more complicated distributions of the SOC fields, where analytic calculations are not available.  

\begin{figure}[htp]
\begin{center}
\includegraphics[scale=0.4]{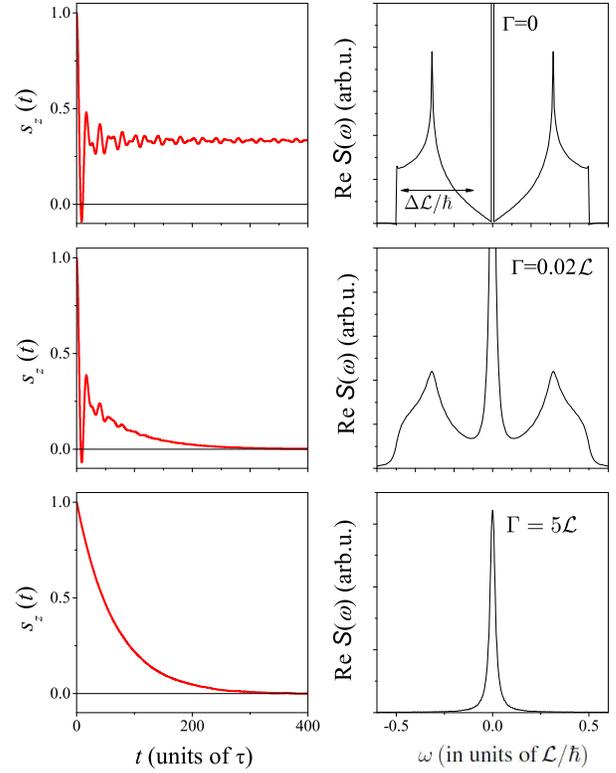}
\caption{The spin dynamics in the presence of the Dresselhaus spin-orbit coupling (from Eq. \eqref{DresselhausHamiltonian} with $\mathcal{L}=1$) for different momentum scattering rates ($\Gamma$). The horizontal arrow indicates the $\Delta\mathcal{L}$ distribution width of the SOC. Note the presence of a Dirac-delta function at $\omega=0$ in $\text{Re}\mathcal{S}(\omega)$ in the clean limit and the gradual broadening with increasing $\Gamma$. The beating pattern in the time domain for $\Gamma=0$ is not noise and is related to the details of the Dresselhaus SOC. The D'yakonov-Perel' limit manifests itself as an exponential time decay in $s_z$and a single Lorentzian at $\omega=0$ for the $\text{Re}\mathcal{S}(\omega)$. The novel regime is identified in the middle graph: an initial rapid decoherence is followed by a longer exponential decay.}
\label{Fig4_Dresselhaus}
\end{center}
\end{figure}

First, we note that the choice of the 2D electron gas with the Bychkov-Rashba SOC is somewhat exceptional as its $\text{Re}\mathcal{S}(\omega)$ in the clean limit 
consists of two delta functions (when quantization is along the $z$ axis), i.e. this model has a zero width of the SOC field distribution. Generally, the width of the SOC field distribution ($\Delta \mathcal{L}=\hbar\Delta \Omega$) could be sizeable, i.e. comparable to the average SOC field ($\left<\mathcal{L}\right>=\hbar\left<\Omega\right>$). In fact, the Dresselhaus SOC represents such a case. Its Larmor frequency distribution reads: 
\begin{equation}
\bm{\Omega}(\bm{k}) =\frac{\mathcal{L}}{\hbar k^3_{\text{F}}}\left[k_x\left(k_y^2-k_z^2\right),k_y\left(k_z^2-k_x^2\right),k_z\left(k_x^2-k_y^2\right)\right].
\label{DresselhausHamiltonian}
\end{equation}
The corresponding $\text{Re}\mathcal{S}(\omega)$ in the clean limit is shown in Fig. \ref{Fig4_Dresselhaus} obtained with $\mathcal{L}=1$. The SOC field distribution is sizeable and $\Delta \mathcal{L}$ and $\left<\mathcal{L}\right>$ have the same order of magnitude.

Two features are observed for this type of SOC in the clean limit: a Dirac-delta function in $\text{Re}\mathcal{S}(\omega=0)$ and "beating" in $s_z(t)$. While similar in its form, the $\text{Re}\mathcal{S}(\omega)$ function is not identical to the histogram of the $\left| \bm{\Omega}(\bm{k}) \right|$ Larmor frequency distribution, which is evident by the presence of the Dirac-delta function in the earlier. This is due to some geometric factors which appear in the calculation of $\text{Re}\mathcal{S}(\omega)$ and is discussed in depth in the Supplementary Material. The beats in the time domain data are the consequence of coherent spin oscillations and its details are specific for the angular distribution of the SOC field. 

Fig. \ref{Fig4_Dresselhaus}. also presents the time dependence of $s_z$ and $\text{Re}\mathcal{S}(\omega)$ for finite $\Gamma$. The D'yakonov-Perel' limit is recovered when $\Gamma$ is much larger than $\left<\mathcal{L}\right>$ and $\Delta\mathcal{L}$. However, a novel regime is identified when $\Gamma \lesssim \left(\left<\mathcal{L}\right>,\Delta\mathcal{L}\right)$. Then, an initial rapid dephasing due to the distribution of SOC fields is present in agreement with Ref. \onlinecite{FabianRMP}, however the dephasing is not complete (i.e. $s_z \neq 0$) and the remaining ensemble $s_z$ decays on the timescale of $\tau=\hbar/\Gamma$ only. This feature is due to the presence of the Dirac-delta function in addition to the SOC fields at $\omega \neq 0$.
This observation mimics the situation encountered in pulsed NMR spectroscopy\cite{SlichterBook}: therein a rapid dephasing (on a timescale denoted as $T_2^*$) is caused by local magnetic field inhomogeneities which is not accompanied by a true information loss. It is followed by a true relaxation (a timescale denoted as $T_2$or in the absence of an external field $T_1=T_2$) where the information is inevitably lost. This phenomenon leads to the presence of NMR spin-echo, i.e. the spins can be restored in-phase on a timescale within $T_2$ with a suitable external excitation. Our observation predicts that a similar scheme may lead to the observation of spin-echo in semiconductors under the circumstances which correspond to the situation shown in Fig. \ref{Fig4_Dresselhaus}. 

\begin{table*}[t]
\begin{center}
\caption{Summary of the relaxation regimes which are encountered for different values of the momentum scattering rate $\Gamma$, average SOC energy $\left<\mathcal{L}\right>$ and its spread $\Delta \mathcal{L}$.}
\begin{tabular}{ l c }
\hline
  Condition & Relaxation type \\
	\hline
  $\Gamma \gg \left(\left< \mathcal{L} \right>,\Delta\mathcal{L} \right)$ & exponential (DP regime), $\Gamma_{\text{s}}=\alpha\mathcal{L}^2/\Gamma$ \\
$\Delta \mathcal{L} \ll \Gamma \lesssim \left< \mathcal{L}\right>$ & oscillatory+exponential decay, $\Gamma_{\text{s}}=\alpha\Gamma$\\
	$\Gamma \lesssim \left(\left< \mathcal{L} \right>,\Delta\mathcal{L} \right)$ & non-exponential, $T_2^*=\hbar/\Delta\mathcal{L}$, $T_{1,2}=\hbar/\Gamma$\\
	\hline

	\label{Summary_table}
\end{tabular}
\end{center}

\end{table*}

We believe that the well-know Dresselhaus Hamiltonian is general enough to properly capture the essential features of spin-relaxation for the entire $\left(\Gamma,\left< \mathcal{L} \right>,\Delta \mathcal{L}\right)$ phase space. It is also the most important Hamiltonian which is relevant for most III-V semiconductors where bulk spatial inversion symmetry breaking occurs. We summarize our qualitative findings in a compact form in Table \ref{Summary_table}., which is briefly repeated herein: 
i) the conventional DP regime occurs when $\Gamma$ is much larger than the SOC, ii) the spin decay is oscillatory and spin-relaxation time has the same order of magnitude as momentum relaxation time when the SOC is significant but its distribution is sharp: $\Delta \mathcal{L} \ll \Gamma \lesssim \left< \mathcal{L}\right>$, iii) an NMR-like rapid spin dephasing followed by a true spin-relaxation occurs when $\Gamma \lesssim \left(\left< \mathcal{L} \right>,\Delta\mathcal{L} \right)$.

We finally comment on the most important approximation of our approach, i.e. that the SOC is smaller than the kinetic energy. This approximation is valid for most technically relevant semiconductors but the opposite is true for heavy elements such as e.g. Bi, where the SOC becomes the leading energy term. Those materials, however, are characterized by very unconventional spin-dynamical properties whose study was attempted in e.g. Ref. \onlinecite{DoraSimonTopIns}.

\section*{Conclusions}
In conclusion, we studied spin-relaxation in materials without a spatial inversion symmetry considering the full phases space of the momentum scattering, the SOC field strength and its distribution properties. Our main tool was a Monte Carlo method which conceptually follows the D'yakonov-Perel' relaxation mechanism. The method is validated by a comparison to a diagrammatic approach based calculation of the spin-relaxation parameters. We identify a compelling analogy between spin-relaxation and the textbook description of NMR spectroscopy in the presence of motion. The method, when applied for the Dresselhaus SOC Hamiltonian, allowed to expand our knowledge about the different spin-relaxation regimes.

\section*{Acknowledgements}
Work supported by the ERC Grant Nr. ERC-259374-Sylo and by the Hungarian National Research, Development and Innovation Office (NKFIH) Grant Nr. K119442. A.K. acknowledges the Bolyai Program of the Hungarian Academy of Sciences.

\section*{Author Contributions}
LS and BD performed the analytic calculations, LS made the numerical studies. AK validated the results. FS outlined the overall composition of the paper and all authors contributed to writing the manuscript.

\section*{Additional Information}
\textbf{Competing financial interests:} The authors declare no competing financial interests.

\section*{Note added}
We became aware of a contribution\cite{Cosacchi}, which investigates similar phenomena such as our work, shortly prior to submitting our manuscript.

\appendix

\newpage
\clearpage

\sloppy

\section{The effect of the $\mathcal{O}\left(\mathcal{L}/E_{\text{F}}\right)$ approximation}

\begin{figure}[htp]
\begin{center}
\includegraphics[scale=0.3]{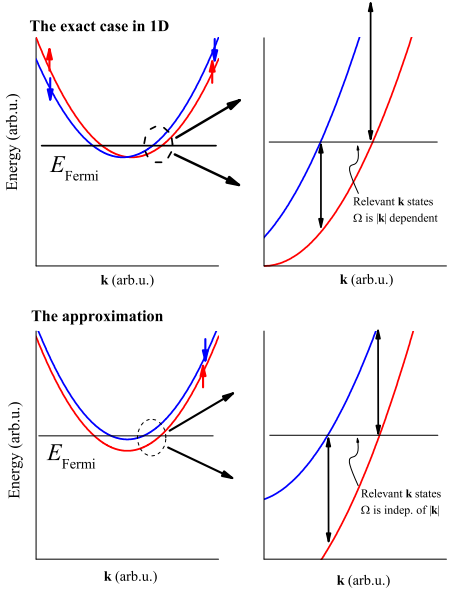}
\caption{Schematics of the $\mathcal{O}\left(\mathcal{L}/E_{\text{F}}\right)$ approximation for a one-dimensional quadratic band dispersion. For such a dispersion, the Bychkov-Rashba type SOC shifts the up/down dispersions to the right/left, respectively. At the Fermi surface intersection, the SOC induced splitting is $\left|\bf{k}\right|$ dependent. In contrast, our approximation is equivalent to substituting it by a hypothetical, Zeeman-like split band structure, where the SOC induced splitting is independent of $\left|\bf{k}\right|$.}
\label{FigS1_approximation}
\end{center}
\end{figure}

It was mentioned in the main text that both the Monte Carlo and the diagrammatic technique in 
Ref. \onlinecite{BurkovBalents} neglects the effect of the SOC on the Fermi surface. 
A proper calculation should consider that the bands are split due to the SOC and therefore 
the corresponding $\Omega(\bf{k})$ is $\left|\bf{k}\right|$ dependent due to this effect. In contrast, our approximation neglects this effect and Fig. \ref{FigS1_approximation}
 depicts this approximation for a one-dimensional band dispersion. The figure suggests that corrections to our approximation are on the order of $\mathcal{O}\left(\mathcal{L}/E_{\text{F}}\right)$.

\section{Relationship between $\mathcal{S}$ and the dynamic spin-susceptibility}

\newcommand{\LRashba}{\mathcal{L}_\mathrm{R}}
\newcommand{\LDressel}{\mathcal{L}_\mathrm{D}}

We discussed in the main text that the time decay of a spin-polarized ensemble (described by $s_z(t)$) is calculated with a Monte Carlo approach for both the clean and
 dirty cases. We found that the real part of its Fourier transform, $\text{Re}\mathcal{S}(\omega)$ can be conveniently displayed in order to demonstrate the spin-relaxation properties. 
We also showed that a relation between $\text{Re}\mathcal{S}(\omega)$ and the dynamic spin-susceptibility, $\chi(\omega)$ holds:
\begin{gather}
\omega \cdot \text{Re}\mathcal{S}(\omega) \propto \text{Im}\chi(\omega).
\end{gather}

\begin{figure}[htp]
\begin{center}
\includegraphics[scale=0.3]{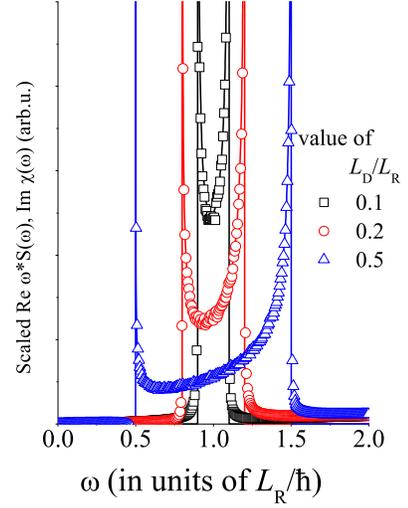}
\caption{Comparison of $\omega\cdot \mathcal{S}(\omega)$ as obtained from our Monte Carlo results (symbols) and the analytic calculations of $\text{Im}\chi(\omega)$ in Ref. \onlinecite{ErlingssonPRB} for various parameters of the SOC Hamiltonian. Note the agreement between the two kinds of data besides a vertical scaling factor.}
\label{FigS2_BR_Dress_susc}
\end{center}
\end{figure}

We present the above relationship on a quantitative agreement which we numerically obtained by comparing our Monte Carlo results on $\mathcal{S}(\omega)$ with analytic calculations 
of $\chi(\omega)$ for a particular Hamiltonian which is available in the literature (Ref. \onlinecite{ErlingssonPRB}). The result is shown in Fig. \ref{FigS2_BR_Dress_susc} with the calculation details as follows. 

Erlingsson \textit{et. al.} (Ref. \onlinecite{ErlingssonPRB}) calculated $\chi(\omega)$ for a two-dimensional electron gas for a Hamiltonian containing both Bychkov-Rashba and Dresselhaus type SOC terms as follows:

\begin{equation}
  H_0 = \frac{\hbar^2\bm{k}^2}{2m}
           + \frac{\LRashba}{k_\mathrm{F}}(s_x k_y - s_y k_x)
           + \frac{\LDressel}{k_\mathrm{F}}(s_y k_y - s_x k_x).
					\label{BR_D_Hamiltonian}
\end{equation}
where $\LRashba$ and $\LDressel$ are the strengths of the Bychkov-Rashba and the Dresselhaus type spin-orbit couplings, respectively.

The dynamic spin-susceptibility was calculated in the absence of momentum relaxation and when $\LRashba \sim \LDressel \ll E_\mathrm{F}$. 
This limit matches the above discussed $\mathcal{O}\left(\mathcal{L}/E_{\text{F}}\right)$ approximation. Eq.~15 in Ref. \onlinecite{ErlingssonPRB} gives the dynamic spin-susceptibility for the $x$ direction as:

\begin{equation}
  \begin{aligned}
    &\chi_{xx}(\omega) = \lim_{\eta\rightarrow 0+}\frac{m}{2\pi\hbar^2}
      \Bigg(
        1
        +\\ 
        &\frac {
            \left( \omega + \mathrm{i} \eta \right)^2 
          }{
            \sqrt{
              \left(
                \frac{
                  \LRashba + \LDressel
                }{
                  \hbar
                }
              \right)^2
              - \left( \omega + \mathrm{i} \eta \right)^2 
            }
            \sqrt{
              \left(
                \frac{
                  \LRashba - \LDressel
                }{
                  \hbar
                }
              \right)^2
              - \left( \omega + \mathrm{i} \eta \right)^2 
            }
          }
      \Bigg)
	\label{ErlingssonResult}		
  \end{aligned}
\end{equation}

Fig. \ref{FigS2_BR_Dress_susc}. demonstrates a good agreement between the two kinds of data which supports the statement in the main text concerning the connection between $\mathcal{S}$ and $\chi$.

\section{ Analytic treatment of the spin dynamics by the time evolution in the clean limit }

We give an analytic description of the time evolution of the magnetization of an electron ensemble subjected to internal SOC fields to verify further the Monte Carlo method applied in the main text since this description is equivalent to the numerical method in the clean limit.
We consider the SOC in the form $H_{\rm 0} = \frac{1}{2}\mathbf{\Omega}(\mathbf{k})\cdot  {\bm{\sigma}}$, where $ {\bm{\sigma}}$ is a vector composed by the Pauli matrices,  and $\mathbf{\Omega}(\mathbf{k})$ is the $\mathbf{k}$-dependent internal SOC field. 

The time evolution of the state of an electron under the SOC is determined by the time evolution operator $U(t)={\rm exp}(-iH_{\rm 0}t/\hbar)$.
Supposing that the electron is initially in spin-up state, i.e. its spin is polarized along the $z$ direction,
\begin{eqnarray}
| \psi(0) \rangle = | \uparrow \rangle =   \frac{ v_2^{(-)} | + \rangle - v_2^{(+)} | - \rangle }{v_2^{(-)} v_1^{(+)}+v_2^{(+)}v_1^{(-)}} , 
\end{eqnarray}   
its state ket at time $t$ is obtained as
\begin{eqnarray}
| \psi(t) \rangle =  \frac{ v_2^{(-)} {\rm e}^{-i E_{+}t/\hbar} | + \rangle - v_2^{(+)} {\rm e}^{-i E_{-}t/\hbar} | - \rangle }{v_2^{(-)} v_1^{(+)}+v_2^{(+)}v_1^{(-)}} \label{eq-psit}
\end{eqnarray}   
by applying the time evolution operator.
Here, $| \pm \rangle = [v_1^{(\pm)}, v_2^{(\pm)}]$ and $E_{\pm}$ are the eigenkets and eigenenergies of the Hamiltonian $H_{\rm 0}$.
Finally, the time development of the $z$ component of the electron spin is obtained as
\begin{eqnarray}
 S_{z} (t, \mathbf{k}) &=& \langle \psi(t) | \hat{S}_{z} | \psi(t) \rangle \nonumber\\
 &=& \frac{ \Omega_x({\mathbf k})^2 + \Omega_y({\mathbf k})^2}{ \Omega({\mathbf k})^2 }  \cos ( \Omega({\mathbf k}) t )  + \frac{\Omega_z({\mathbf k})^2}{\Omega({\mathbf k})^2 } ,
 \nonumber\\
\label{eq-expSz1}
\end{eqnarray}
where $\Omega = \sqrt{\Omega_x^2 + \Omega_y^2 + \Omega_z^2}$.
The quantity $S_{z}(t)$, i.e. the $z$ component of a spin ensemble, calculated by Monte Carlo method in the main text is obtained as
\begin{eqnarray}
 S_{z} (t) 
 = \int_{\rm F.S} d\mathbf{k} S_{z} (t, \mathbf{k})
 \label{eq-expSz2}
\end{eqnarray}
within this approach, i.e. by integration over $\mathbf{k}$  on the Fermi surface.
Arbitrary $ \mathbf{\Omega}({\mathbf k})$ can be considered in the Hamiltonian $H_{\rm 0}$ such as the two-dimensional Bychkov-Rashba SOC or the three-dimensional Dresselhaus case discussed in the main text. 
We note that for a complicated distribution of the SOC fields the $\mathbf{k}$ integration in Eq.~(\ref{eq-expSz2}) might not be performed analytically. In such a case, the integration can be performed by choosing random $\mathbf{k}$ values on the Fermi surface.

When this calculation is performed according to Eq. \eqref{eq-expSz1} for different model Hamiltonians such as those given in the main text, we obtained numerically identical results (data not shown) as for the Monte Carlo, and the analytic result for the dynamic spin susceptibility given in Eq. \eqref{ErlingssonResult} is reproduced as well.

From expression~(\ref{eq-expSz1}) it is obvious that in a two-dimensional case, i.e. when $\Omega_{z}=0$:
\begin{eqnarray}
 S_{z} (t, \mathbf{k}) 
 = \cos ( \Omega({\mathbf k}) t ).
\label{eq-expSz3}
\end{eqnarray}
By taking the Bychkov-Rashba SOC, $\mathbf{\Omega}(\mathbf{k}) = \frac{{\cal L}}{\hbar k_{\rm F}}[k_x,k_y,0]$,  $\Omega({\mathbf k}) $ becomes $\mathbf{k}$ independent as $\Omega({\mathbf k}) = {\cal L}/\hbar = \Omega$, which means a single oscillating component in $S_{z}(t)$ as it is shown by dashed curve in Fig.2 of the main text, and two Dirac-delta peaks at $\pm \Omega$ in the real part of the Fourier transform $\mathcal{S}(\omega)$.
In three-dimensional cases, there is always a $t$-independent non-zero term in $S_z(t)$ coming from the last term  in Eq.~(\ref{eq-expSz1}), which explains the finite $S_z$ value in Fig.4 of the main text with $\Gamma=0$. This non-zero and time independent term corresponds to a Dirac-delta function centered on $\omega=0$ in $\mathcal{S}(\omega)$.

\begin{figure}[htp]
\begin{center}
\includegraphics[scale=0.4]{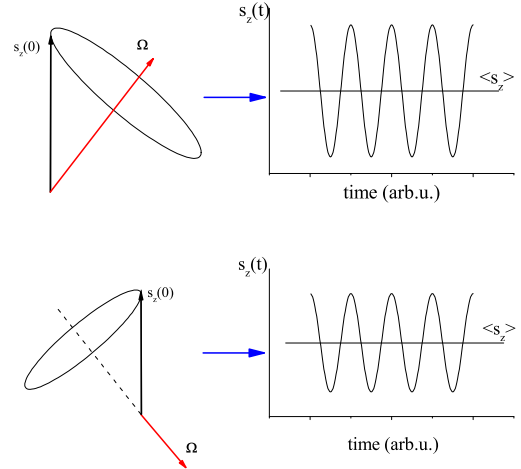}
\caption{Schematic depiction of the spin precession around the SOC fields and the corresponding $\mathbf{\Omega}(\mathbf{k})$ vectors when the spins start along the $z$ direction at $t=0$. Note that for an arbitrary $\mathbf{\Omega}(\mathbf{k})$ which is not in the $x-y$ plane, the precession retains a finite \emph{positive} $s_z$ value.}
\label{FigS3_Dresselhaus_dirac_delta}
\end{center}
\end{figure}

The origin of this effect is further supported by a geometric consideration which is depicted in Fig. \ref{FigS3_Dresselhaus_dirac_delta}. The presence of the Dirac-delta peak for $\mathcal{S}(\omega=0)$ is a generic feature and its absence for the two-dimensional electron gas and the Bychkov-Rashba SOC is an exeption. For the latter, when the spins are aligned perpendicular to the 2D plane, all SOC fields are in the plane, i.e. the precession of the spins around the built in $\mathbf{\Omega}(\mathbf{k})$ results in a zero-averaged net magnetization. However, for a general distribution of the SOC fields and the corresponding $\mathbf{\Omega}(\mathbf{k})$ vectors, the precession of the spins retains a finite positive $s_z$ component as Fig. \ref{FigS3_Dresselhaus_dirac_delta}. depicts. A straightforward geometric consideration shows that the $\left<s_z\right>$, i.e. the Dirac delta function strength is given by $\Omega_z^2/\Omega^2$ ($\Omega_z$ and $\Omega$ are the $z$ component and the magnitude of the $\mathbf{\Omega}(\mathbf{k})$ vector, respectively) for a particular $\mathbf{\Omega}(\mathbf{k})$ component. Similarly, we obtain that the amplitude of the oscillation goes as $1-\Omega_z^2/\Omega^2$, in full agreement with Eq. \eqref{eq-expSz1}.

\begin{figure}[htp]
\begin{center}
\includegraphics[scale=0.35]{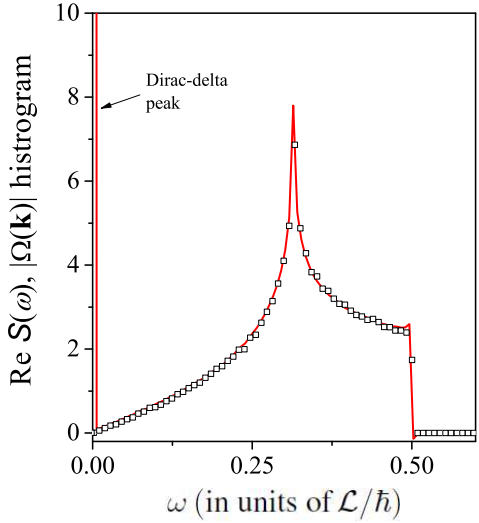}
\caption{Comparison of the Monte Carlo based $\text{Re}\mathcal{S}(\omega)$ (solid line) and the histogram of the $\left|\Omega(\mathbf{k})\right|$ (open symbols). Note that there is no scaling parameters between the two kinds of data.}
\label{FigS4_Dresselhaus_distribution}
\end{center}
\end{figure}

In Fig. \ref{FigS4_Dresselhaus_distribution}., we present the comparison between the $\text{Re}\mathcal{S}(\omega)$ and the histogram of the internal Larmor frequency distribution for the three-dimensional Dresselhaus model. The latter data integrated for the positive frequencies gives the average of the SOC fields which coincides with the frequency value where it is peaked and its weighted integral, i.e the integral of the histogram values multiplied by the related frequency, gives unity since all weights are summed up in this way.

\section{Analogy between the motional narrowing and the spin-relaxation for the 2D BR modell}

\begin{figure}[htp]
\begin{center}
\includegraphics[scale=0.3]{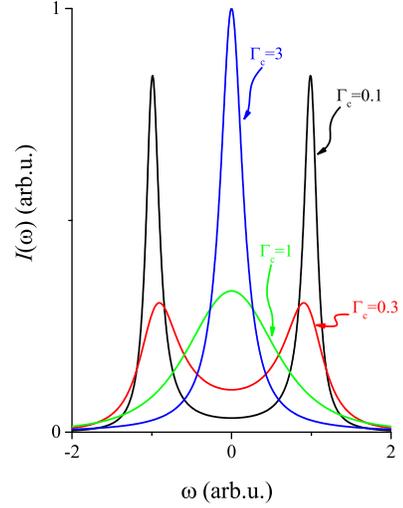}
\caption{Simulated lineshapes for the two-site NMR motional narrowing problem. Note that two peaks are observed for smaller values of $\Gamma$ whose linewidth increases with increasing $\Gamma$. In contrast, a single, motionally narrowed peak is observed for larger $\Gamma$ values, whose linewidth \emph{decreases} with increasing $\Gamma$.}
\label{FigS5_motional_narrowing}
\end{center}
\end{figure}

Abragam \cite{AbragamBook} considered the so-called two-site NMR motional narrowing problem: a nuclei is allowed to jump with the transition rate $\Gamma_{\text{c}}=1/\tau_{\text{c}}$ between two 
sites with different local Larmor frequencies: $\pm \Omega$ around a central Larmor frequency (defined as zero in this case). The resulting NMR lineshape is shown in Fig.~\ref{FigS5_motional_narrowing} 
for a fixed $\Omega=1$ and different values of the jumping frequency, $\Gamma_{\text{c}}$. The analogy between the spin-relaxation and the motional narrowing is clear: the $\pm \Omega$ local Larmor 
frequencies correspond to the built-in Zeeman field distribution of the spin-relaxation problem and the jumping frequency ($\Gamma_{\text{c}}$) of the motional narrowing problem corresponds to the 
$\Gamma$ momentum relaxation rate (besides a factor 2 which is discussed below). The analogy can be quantified for the simplest case as follows.


\begin{gather}
  I(\omega) = \Re \frac{ 2\mathrm{i}\omega
                         + 4\Gamma_{\text{c}}
                       }{ \left( \Omega^2-\omega^2 \right)
                          + 2\mathrm{i}\omega\Gamma_{\text{c}}
                       }.
  \label{Abragam_formula1}
\end{gather}

The denominator of Eq.~\eqref{Abragam_formula1} has poles at $\omega_{1,2}=\mathrm{i}\left(\Gamma_{\text{c}}\pm \sqrt{\Gamma_\text{c}^2-\Omega^2} \right)$, i.e. Eq. \eqref{Abragam_formula1} can be rewritten as:

\begin{gather}
  I(\omega) = \Re \left( \frac{A}{\omega-\omega_1}
                         + \frac{B}{\omega-\omega_2}
                  \right),
  \label{Abragam_formula2}
\end{gather}

\noindent where $A=-\mathrm{i}\frac{\Gamma_{\text{c}}+\Delta}{\Delta}$ and $B=\mathrm{i}\frac{\Gamma_{\text{c}}-\Delta}{\Delta}$. Herein, we introduced $\Delta=\sqrt{\Gamma_{\text{c}}^2-\Omega^2}$. Evaluation of Eqs. \eqref{Abragam_formula1} and \eqref{Abragam_formula2} yields the curves shown in Fig. \ref{FigS5_motional_narrowing}.

Our definition of $\Gamma$ differs from that of $\Gamma_{\text{c}}$ in Abragam's work in a factor 2 as discussed herein, as $\Gamma$ corresponds to the momentum relaxation rate and $\Gamma_c$ corresponds to the transition rate between the two states. The rate equations for the two state's populations:

\begin{equation}
  \begin{aligned}
    \dot{n}_1 &= -\frac1{\tau_c}\left( n_1 - n_2 \right) \\
    \dot{n}_2 &=  \frac1{\tau_c}\left( n_1 - n_2 \right) \\
    \partial_t(n_1-n_2) &= -\frac2{\tau_c}\left( n_1 - n_2 \right).
  \end{aligned}
\end{equation}

\begin{figure}[htp]
\begin{center}
\includegraphics[scale=0.3]{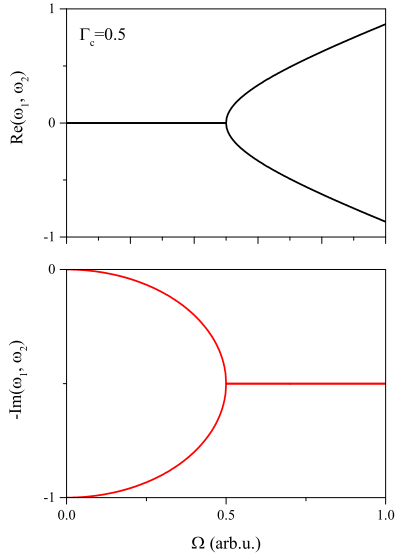}
\caption{Real and imaginary parts of the $\omega_{1,2}$ roots as defined above from the NMR motional narrowing problem.}
\label{FigS6_motional_narrowing_roots}
\end{center}
\end{figure}

We can see that $\frac1{\tau_\text{m}}=\frac2{\tau_c}$, so $\Gamma=2\Gamma_c$.
With this change, the real and imaginary values of the above defined $\omega_{1,2}$ are shown in Fig. \ref{FigS6_motional_narrowing_roots}. 
The real part of the roots describe the position of the two peaks and the imaginary parts describe the linewidths in agreement with Fig. \ref{FigS6_motional_narrowing_roots}. Remarkably, this figure is \emph{identical} to the spin-relaxation problem for the 2D Rashba model in the main paper with a straightforward identification of the correspondence of the two parameters.

\bibliographystyle{naturemag}
\bibliography{Tubes2011June_new}

\newpage

\end{document}